\documentclass[11pt]{article}
\pdfoutput=1 

\usepackage{jheppub}

\usepackage{color}
\usepackage{amsmath}
\usepackage{verbatim}   
\usepackage{subfigure}  
\usepackage{acronym}

\usepackage{amsfonts}
\usepackage{amssymb}
\usepackage{mathrsfs}
\usepackage{graphicx}
\usepackage{multirow}
 \usepackage{slashed}
 \usepackage{epsfig,multicol,bbm}
 \usepackage{url}

\newcommand{\newc}{\newcommand}
\newc{\gsim}{\lower.7ex\hbox{$\;\stackrel{\textstyle>}{\sim}\;$}}
\newc{\lsim}{\lower.7ex\hbox{$\;\stackrel{\textstyle<}{\sim}\;$}}
\newc{\gev}{\,{\rm GeV}}
\newc{\mev}{\,{\rm MeV}}
\newc{\ev}{\,{\rm eV}}
\newc{\kev}{\,{\rm keV}}
\newc{\tev}{\,{\rm TeV}}

\def\beq{\begin{equation}}
\def\eeq{\end{equation}}
\def\bea{\begin{eqnarray}}
\def\eea{\end{eqnarray}}
\def\bitem{\begin{itemize}}
\def\eitem{\end{itemize}}
\newcommand{\bec}{\begin{center}}
\newcommand{\eec}{\end{center}}

\newcommand{\mb}[1]{\boldsymbol{#1}}
 \newcommand{\MeV}{{\mathrm {MeV}}}
  \newcommand{\GeV}{{\mathrm {GeV}}}
   
      \newcommand{\MP}{M_{\mathrm {Pl}}}

\def\bar#1{\overline{#1}}

\def\inv{^{\raise.15ex\hbox{${\scriptscriptstyle -}$}\kern-.05em 1}}
 
\def\lbar{{\lower.35ex\hbox{$\mathchar'26$}\mkern-10mu\lambda}}

\let\<=\langle
\let\>=\rangle

\let\+=\uparrow

\def\OO{\mathcal{O}}

\begin{document}

\hfill \vspace{-5mm} OUTP-12-21P

\title{\Huge Exodus\\ \vspace{3mm} {\LARGE Hidden origin of dark matter and baryons}}

\author{James Unwin}
\emailAdd{unwin@maths.ox.ac.uk}
\affiliation{Rudolf Peierls Centre for Theoretical Physics,
University of Oxford,\\
1 Keble Road, Oxford,
OX1 3NP, UK}
\affiliation{Mathematical Institute,
University of Oxford,
24-29 St Giles, Oxford,
OX1 3LB, UK}

\date{\today}

\abstract{We propose a new framework for explaining the proximity of the baryon and dark matter relic densities $\Omega_{\rm{DM}}\approx5\Omega_B$. The scenario assumes that the number density of the observed dark matter states is generated due to decays from a second hidden sector which simultaneously generates the baryon asymmetry. In contrast to asymmetric dark matter models, the dark matter can be a real scalar or Majorana fermion and thus presents distinct phenomenology.  We discuss aspects of model building and general constraints in this framework.  Moreover, we argue that this scenario circumvents several of the experimental bounds which significantly constrain typical models of asymmetric dark matter. We present a simple supersymmetric implementation of this mechanism and show that it can be used to obtain the correct dark matter relic density for a bino LSP.}

\maketitle


\section{Introduction}

The matter of the observable universe, the {\em visible sector}, exhibits a rich structure of interacting states and, a priori, one might expect that the dark matter (DM) can be part of an equally complicated (set of) hidden sector(s). Such a suggestion is not implausible from a UV perspective as, for example, it has been argued that the topological complexity of generic string theory compactifications result in multiple hidden sectors sequestered from the visible sector \cite{seq}. In this context it is quite conceivable that exchanges between dark and visible sectors can have cosmological consequences (beyond the usual moduli problem) and this reasoning motivates us to propose a new mechanism for cogenerating the DM relic density  $\Omega_{\mathrm{DM}}$ and baryon density $\Omega_B$ in such a way that the close coincidence $\Omega_{\mathrm{DM}}\approx5\Omega_B$ is explained without necessarily resorting to the assumption that that the DM itself carries a particle-antiparticle asymmetry.

\begin{figure}[t!]
\begin{center}
\includegraphics[height=59mm]{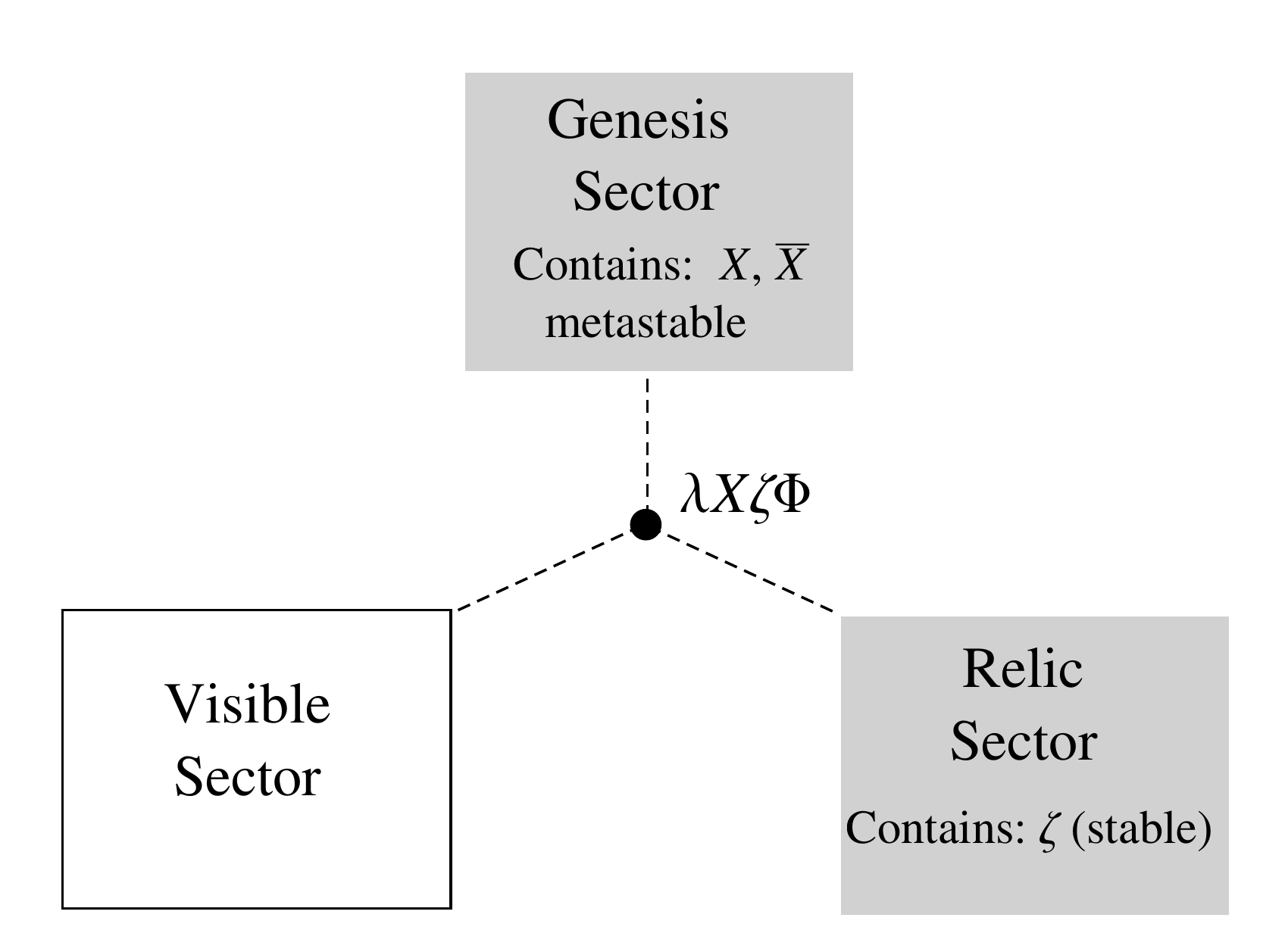}
\label{Fig1}
\caption{Illustration of the three sectors connected via a weak trisector interaction $\lambda X\zeta\Phi$, where $\Phi$ is some Standard Model singlet operator involving Standard Model states which violates $B-L$.}
\end{center}
\end{figure}

Hidden sectors can readily accommodate large CP violation which can lead to asymmetries in hidden sector states, for instance, via analogues of traditional baryogenesis mechanisms. If this asymmetry is transferred to the visible sector then it can generate the baryon asymmetry and thus set the present day baryon density \cite{ADMgenesis}. In the new framework presented here the interplay between two hidden sectors, connected via a weak trisector coupling involving the visible sector, is used to explain the coincidence of cosmological densities $\Omega_{\mathrm{DM}}\approx5\Omega_B$. An asymmetry in some approximately conserved quantum number is generated in a hidden {\em genesis sector} resulting in an asymmetry between a state $ {X}$ and its anti-partner $\bar X$. The genesis sector then evolves such that the symmetric component of $X$ annihilates away and the abundance of $X$ is set by the asymmetry. Subsequently, the asymmetric component of $ {X}$ decays to the visible sector and a second hidden sector, the {\em relic sector}, in a manner that cogenerates the DM and the baryon asymmetry. This scenario is illustrated schematically in Fig.~\ref{Fig1}. Since the mechanism proposed here occurs after the genesis of a hidden sector asymmetry, and the DM and baryon asymmetry are due to energy leaving the genesis sector, we refer to this mechanism as {\em exodus}.

The mechanism presented here is related to models of asymmetric DM  \cite{ADMgenesis,ADM,Hall:2010jx,Davoudiasl:2010am} and shares some features in common with Hylogenesis \cite{Davoudiasl:2010am}, DM assimilation \cite{D'Eramo:2011ec}, and other scenarios involving  meta-stable states decaying to DM and baryons \cite{others} or multiple sectors \cite{Fischler:2010nk}. However, the construction and phenomenology of the exodus framework is quite distinct, and, notably, unlike existing models our mechanism {\em allows the generation of self-adjoint or asymmetric DM}, whilst simultaneously explaining the coincidence $\Omega_{\rm{DM}}\approx5\Omega_B$. In particular, we shall argue that this framework provides a new possibility for obtaining the correct DM relic density composed of (nearly) pure bino LSP, a scenario which is generally not viable in the conventional freeze-out picture.

This paper is structured as follows, we first provide an overview of the mechanism and discuss the general requirements for connecting $\Omega_{\rm{DM}}$ and $\Omega_B$ in this model. Subsequently, we present a specific example in which we use this mechanism to obtain the correct DM relic density for a bino LSP in Sect.~\ref{Sec3}. We then generalise this specific implementation to explore some general aspects of model building and, outline some variant models involving $L$-violating portal operators and asymmetric DM in Sect.~\ref{Sec4}. Finally, in the concluding remarks we discuss some of the benefits of the exodus mechanism over asymmetric DM models.

\section{Outline of the exodus mechanism}
\label{Sec2}
 
In the framework proposed here the observed DM relic density $\Omega_{\mathrm{DM}}$ is due to an abundance of stable states $ {\zeta}$ in the relic sector. The genesis sector features large C- and CP-violating processes and appropriate out-of-equilibrium dynamics (therefore satisfying the Sakharov conditions \cite{Sakharov:1967dj}) which results in a sizeable asymmetry between the states $X$ and $\bar{X}$, with some approximately conserved quantum number which we denote $\mathscr{X}$.  The state $X$ transforms under the DM stabilising symmetry, but $m_{ {\zeta}}<m_{ {X}}$ and the ${X}$ do not comprise the present day DM relic density. The state $ {X}$ can decay to $ {\zeta}$, however, only via a suppressed intersector interaction, which violates $\mathscr X$ and $B$ (and/or $L$) but conserves some combination of these quantum numbers, e.g.~$B-L+\mathscr X$. For this mechanism to link $\Omega_{\zeta}$ and $\Omega_B$ there are are three general requirements which must be satisfied:
\begin{itemize}
\item The intersector coupling must be sufficiently small that decays of $ {X}$ do not occur until after the symmetric component of $ {X}$ has been removed via pair annihilation. 
\item 
The number density of DM states $n_{ {\zeta}}$ in the relic sector is due only to decays of the residual asymmetric component of $ {X}$ via the trisector coupling, i.e. $n_{ {\zeta}}\big|_{\mathrm{initial}}\simeq0$.
\item 
The relic DM produced via $X$ decays must be entirely responsible for the DM relic density, 
and not  due to thermally produced DM from heating of the relic sector.
\end{itemize}
These conditions ensure that the baryon asymmetry and the DM number density are linked $n_{ {\zeta}}\propto n_{b-\overline{b}}$, as we will discuss in detail below.  We assume that the initial abundance of DM states $ {\zeta}$ is negligible, perhaps due to preferential inflaton decay \cite{Cicoli:2010ha} which results in the relic sector reheating to a temperature lower than $m_\zeta$ (alternative scenarios could be envisaged). This type of cold hidden sector is similar to that considered in models of freeze-in  production of DM \cite{Hall:2010jx,Hall:2009bx}. Importantly the temperature must be significantly lower that $m_\zeta$ to ensure that the tail of the thermal distribution does not populate $\zeta$. For the case of weak-scale $\zeta$ self-interactions this translates into the requirement that $T^{\rm RH}_{\rm relic}\lesssim m_{\zeta}/25$ for $X$ decays to be solely responsible for the DM relic density.

 \begin{figure}[t!]
\begin{center}
\includegraphics[height=65mm]{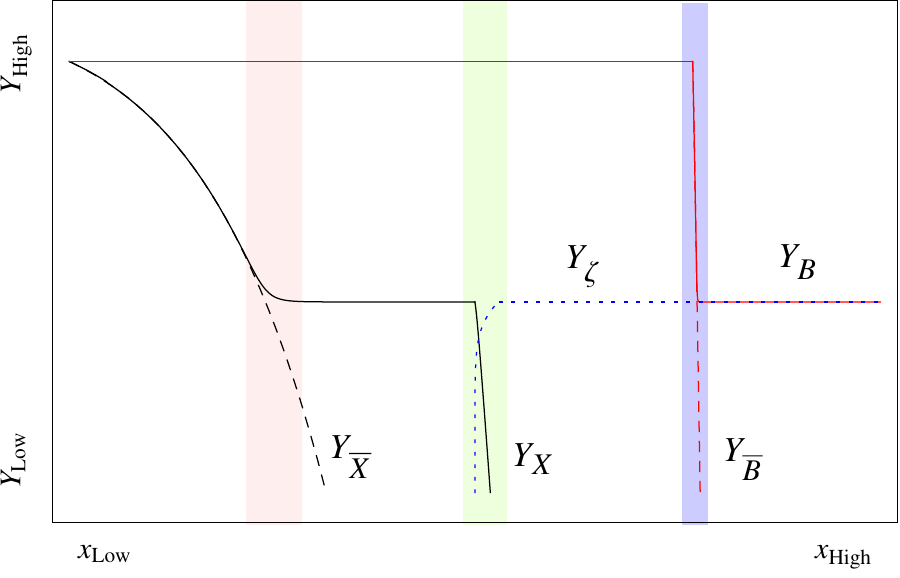}
\caption{Log plot illustrating the cosmological history, it shows (schematically) particle yields $Y\equiv\frac{n}{S}$, where $S$ is the entropy density, against $x\propto T^{-1}$. Black, red and blue solid lines give, respectively, the yields for ${X}$, the baryons and the relic DM $\zeta$ (antiparticles shown dashed). The red shaded region indicates where the $X$ density becomes dominated by the asymmetry, the green region highlights the decays of $X$ and subsequent generation of DM and visible sector asymmetry. In blue is shown the mass threshold of protons past which their number density is Boltzmann suppressed and the baryon density is set by the asymmetry.} 
\label{Fig2}
\end{center}
\end{figure}

Immediately after the $\mathscr X$-genesis in the hidden sector  (shaded red in Fig.~\ref{Fig2}) there is no asymmetry in $B$ or $L$, the number density of $X$ is set by the asymmetry in $\mathscr X$ and $n_{\zeta},n_{\bar X}\approx0 $.  At a later time the $ {X}$ states decay via the trisector coupling producing the DM and an asymmetry in the visible sector which sets the baryon asymmetry (green in Fig.~\ref{Fig2}). As the temperature drops below the baryon mass ($\sim1$ GeV) the baryon number density becomes suppressed and reveals the baryon asymmetry inherited from the genesis sector (blue in Fig.~\ref{Fig2}). We study the Boltzmann equations which describe the asymmetry transfer in the Appendix \ref{App}.

The decay of the asymmetric component of ${X}$ from the genesis sector is solely via the trisector interaction $\lambda {X} {\zeta}\Phi$, for $\Phi$ some Standard Model (SM) singlet operator involving SM states which violates $B-L$. For  preferential inflationary reheating to occur the sectors must be only feebly coupled to each other and consequently, we expect the coefficient of the intersector coupling $\lambda$ to be small. In many realisations the operator   ${X} {\zeta}\Phi$ has high mass dimension and the effective coupling is set by the inverse power of some mass scale $M$, for example  $\frac{1}{M^3} X\zeta u^cd^cd^c$.

 One of the main factors upon which the phenomenology of these models depends is whether the DM particle $\zeta$ carries $\mathscr X$ number. If $\zeta$ does not carry $\mathscr X$ then it can be self-adjoint state (a real scalar or Majorana fermion), thus resulting in symmetric DM. In this case DM annihilations are possible and can lead to indirect detection signals with annihilation profiles. On the other hand, if $\zeta$ is non-self-adjoint (a Dirac fermion or complex scalar), with non-zero $\mathscr X$ number, then these states will inherit the asymmetry of the genesis sector, like the baryons, resulting in asymmetric DM.

\subsection{Lifetime constraints}
\label{3.1}

\begin{figure}[t!]
\begin{center}
\includegraphics[height=47mm]{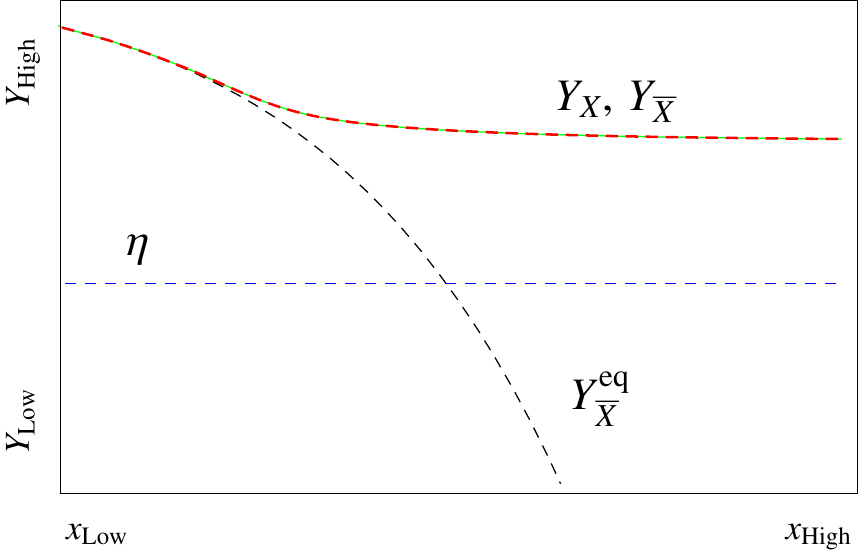}
\hspace{5mm}
\includegraphics[height=47mm]{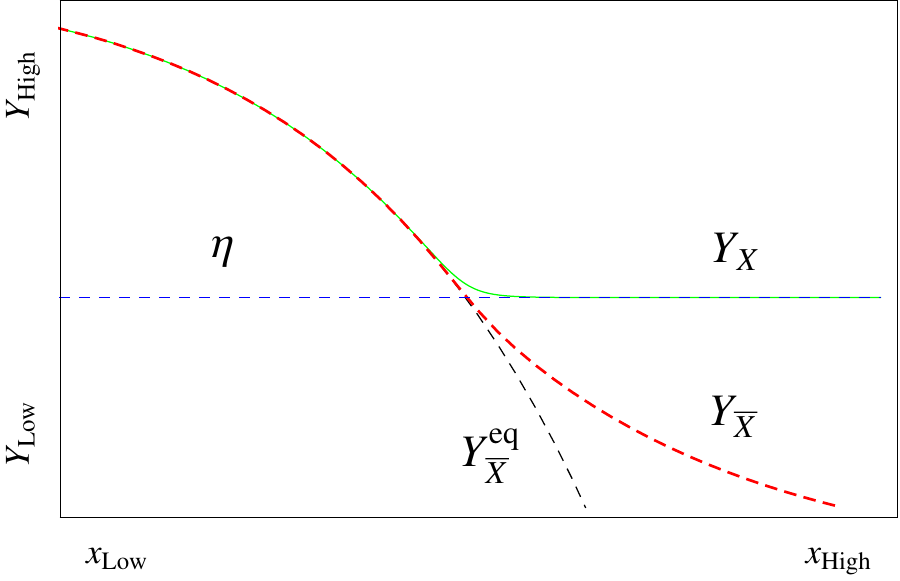}
\caption{Log plot showing (schematically) $Y$ against $x\propto T^{-1}$. The left panel shows the case where freeze-out happens before the critical temperature, consequently, the asymmetry does not set the final density of $ {X}$ and is comparable to the $\bar{ {X}}$ density. Whereas in the right panel, freeze-out happens (just) after the critical temperature and thus $Y_X$ is set by the asymmetry and $Y_{\bar X}\ll Y_X$.} 
\label{Fig3}
\end{center}
\end{figure}

To explain the comparable sizes of the cosmic relic densities $\Omega_\zeta/\Omega_B\approx5$ via the exodus mechanism it is required that the $\mathscr X$ asymmetry alone be responsible for the DM relic density. Consequently we require that the symmetric component annihilates within the lifetime of ${X}$. Here we calculate the minimum lifetime of ${X}$ required such that decays of the $X$ states occur only after the symmetric component is removed via pair annihilation. The $X$ and $\bar X$ could annihilate to either light hidden sector states which do not carry $\mathscr X$, and subsequently redshift away the energy density, or to visible sector states which thermalise before Big Bang nucleosynthesis (BBN). In order for the symmetric component of $X$ to be sufficiently depleted it is necessary that the $X\bar{X}$ annihilations do not freeze-out too quickly, as illustrated in Fig.~\ref{Fig3}.

At high temperatures the $X$ and $\overline{ {X}}$ are in thermal equilibrium with the other states of the genesis sector and when the temperature of the genesis sector $T_{\rm gen}$ drops below the mass of $X$ the number densities of these states decrease exponentially (provided $m_X>|\mu_X|$)
\begin{equation}
\begin{aligned}
n_{X}^{\mathrm{eq}}=\left(\frac{m_XT_{\rm gen}}{2\pi}\right)^{3/2}e^{-\frac{m_X-\mu_X}{T_{\rm gen}}},\hspace{15mm}
n_{\overline{X}}^{\mathrm{eq}}=\left(\frac{m_XT_{\rm gen}}{2\pi}\right)^{3/2}e^{-\frac{m_X+\mu_X}{T_{\rm gen}}}~.
\label{n}
\end{aligned}
\end{equation}
However, once the interaction rate falls below the Hubble rate $H$ the $X\overline{ {X}}$ annihilations freeze-out and their co-moving number densities plateau. The temperature at which annihilations freeze-out $T^{(\rm{FO})}_{\rm gen}$ can be determined by adapting a standard calculation \cite{standard}, which gives the following criteria
\begin{equation}
\begin{aligned}
\frac{m_X}{T^{(\rm{FO})}_{\rm gen}}H\left(T^{(\rm{FO})}_{\rm gen}\right)\sim \left(T^{(\rm{FO})}_{\rm gen}\right)^3\eta_{\mathscr{X}}\sigma ~,
\label{FO}
\end{aligned}
\end{equation}
where $\sigma$ is the annihilation cross section and, introducing $\Delta_{\mathscr{X}}\equiv Y_X -Y_{\overline{X}}$ in terms of the yields $Y_i\equiv n_i/S$, the particle asymmetry is defined $\eta_{\mathscr{X}}\equiv\left(S/n_{\gamma}\right)\Delta_{\mathscr{X}}$, for $S$ the entropy density and $n_{\gamma}$ the photon number density; note $S/n_{\gamma}\approx 7$. For clarity in this section we shall work only in terms of $\eta_{\mathscr{X}}$. Using the standard definition $H(T)=\frac{1.66\sqrt{g_*}T^2}{\MP}$, where $\sqrt{g_*}$ are the number of relativistic degrees of freedom and $\MP=1.2\times10^{19}$ GeV is the Planck mass, we solve eq.~(\ref{FO}) to obtain an expression for the freeze-out temperature
\begin{equation}
T^{(\rm{FO})}_{\rm gen}\sim\sqrt{\frac{1.66\sqrt{g_*}m_X}{M_{\rm{Pl}}\eta_{\mathscr{X}}\sigma}}~.
\label{FOT}
\end{equation}

\begin{figure}[t!]
\begin{center}
\includegraphics[height=65mm]{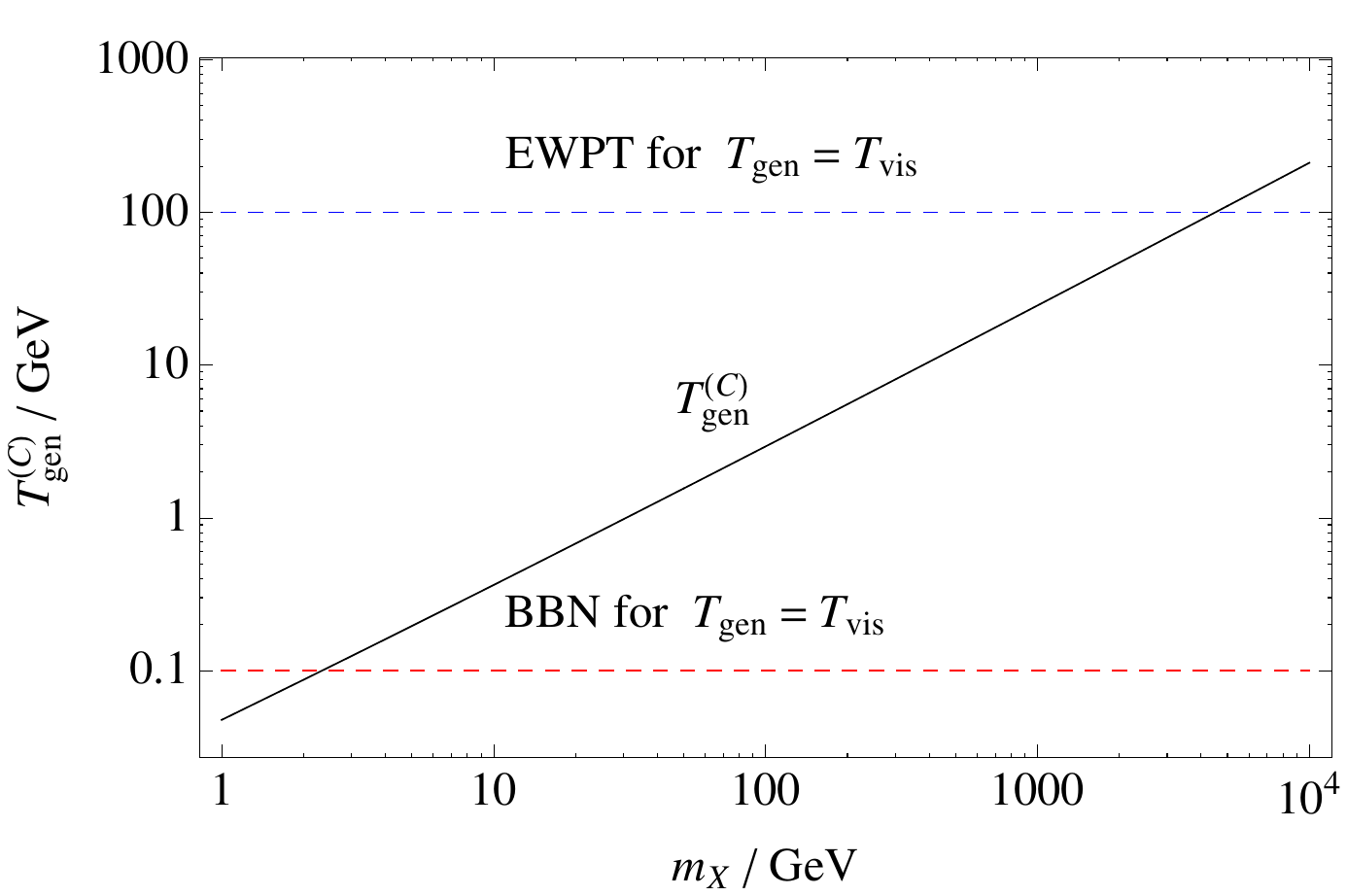}
\caption{The critical temperature $T_{\rm gen}^{(\rm C)}$, below which the symmetric component of $X$ states is sufficiently depleted, for varying $m_X$, with $\eta_{\mathscr{X}}=6.2\times10^{-10}$. In the simplest scenarios the genesis and visible sectors are in thermal equilibrium and we indicate the temperature at BBN (red dashed) and EWPT (green dashed) for this case.} 
\label{FigC}
\end{center}
\end{figure}

Next we calculate the critical temperature $T_{\rm gen}^{(\rm C)}$ for which the number density of $X$ states becomes primarily due to the asymmetry, or equivalently the temperature at which $Y_{\overline{X}}^{\mathrm{eq}}\ll Y_X$. For concreteness we shall insist that $Y_{\overline{X}}^{\mathrm{eq}}<\eta_{\mathscr{X}}/100$. The critical temperature below which this depletion of the symmetric component is achieved is a function of the DM mass. The number density for states with an asymmetry is given by \cite{standard}
\begin{equation} 
n_{\overline{X}}^{\mathrm{eq}}\sim\frac{m_X^3}{\eta_{\mathscr{X}}}e^{-\frac{2m_X}{T_{\rm gen}}}~,
\end{equation}
 and we use $S=\frac{2\pi^2}{45}g_*^ST^3$ to express the depletion condition as follows
\begin{equation}
Y_{\overline{X}}^{\mathrm{eq}}\sim\frac{2\pi^2}{45}\frac{g_*^S\left(T_{\rm gen}^{(\rm C)}\right)^3m_X^3}{\eta_{\mathscr{X}}}e^{-\frac{2m_X}{T_{\rm gen}^{(\rm C)}}}=\frac{\eta_{\mathscr{X}}}{100}~.
\end{equation}
 In Fig.~\ref{FigC} we plot the critical temperature $T_{\rm gen}^{(\rm C)}$ necessary for sufficient annihilation of the symmetric component against $m_X$, taking $\eta_{\mathscr{X}}=6.2\times10^{-10}$, equal to the baryon asymmetry. In the case that the genesis sector is radiation dominated,\footnote{The time-temperature relationship for matter-dominance is of the form $ t\sim H^{-1}\sim \frac{\MP}{\sqrt{\rho}}(1+z)^{-3/2}$ where $z$ is the redshift. For simplicity we shall assume a radiation dominated genesis sector henceforth.}  time can be related to temperature via $ t\sim 1/H(T)\sim \MP/\sqrt{\rho}$ and we can re-express the critical temperature $T_{\rm gen}^{(\rm C)}$ in terms of the minimum time $t_*$ required to adequately deplete the symmetric component
 \beq
t_* \sim \frac{\MP}{1.66\sqrt{g_*}~T_{\rm vis}^2}
 \sim 10^{-6}~{\rm s}~\left(\frac{R}{1}\right)^2~\left(\frac{ {\rm GeV} }{T_{\rm gen}^{(\rm C)}}\right)^2,
 \label{tmin}
 \eeq
where $R=T_{\rm vis}/T_{\rm gen}$ is the quotient of the temperatures of the genesis and visible sector.

In many constructions it is reasonable to assume that in the early universe the genesis sector and visible sector were in thermal equilibrium, before later decoupling. In this case the thermal evolution of the genesis and visible sectors may be very similar and we can compare the critical temperature $T_{\rm gen}^{(\rm C)}$ to important thresholds in the visible sector, such as Big Bang nuclearsynthesis (BBN) and the electroweak phase transition (EWPT), as is shown in Fig.~\ref{FigC}.
 Of course, if the genesis sector and visible sector are not in thermal equilibrium in the early universe, or evolve very differently after decoupling, then the critical temperature can not be readily compared with visible sector milestones.

To demonstrate that viable models can be constructed we must compare the ${X}$  lifetime to the minimum time required to annihilate the symmetric component. For simplicity, let us examine the lifetime of the ${X}$ states, assuming decays via a renormalisable operator with coupling constant $\lambda$ to a two-body final state
\begin{equation}
\begin{aligned}
\tau_X=\Gamma^{-1}_X
\simeq1\times10^{-4}~{\rm s}~\left(\frac{10^{-10}}{\lambda}\right)^2 \left(\frac{10\, \mathrm{GeV}}{m_{ {X}}}\right)~.
\label{tau}
\end{aligned}
\end{equation}
In several scenarios $X$ decays to a multi-body final state via a high dimension operator dressed by the reciprocal of an appropriate scale $M$, this leads to additional phase space suppression of $\Gamma_X$ and the coupling $\lambda$ should be replaced by a ratio of scales, parametrically, $\left(m_X/M\right)^n$ for a $4+n$ dimensional operator.

To construct a proof of principle, let us assume that the temperatures of the visible and genesis sectors are equal and evolve together $T_{\rm{vis}}=T_{\rm{gen}}$. The minimum time required to sufficiently deplete the symmetric component $t_*$ is given by eq.~(\ref{tmin}), with $R=1$. Further, from inspection of Fig.~\ref{FigC}, the critical temperature is $T_{\rm gen}^{(\rm C)}\sim\frac{m_X}{30}$ and thus parametrically
\begin{equation}
t_*\sim10^{-5}\left(\frac{10~\GeV}{m_X}\right)^{2}~{\rm s}~.
\end{equation}
Comparing with eq.~(\ref{tau}) we conclude that viable models, with $\tau_X>t_{*}$, can be constructed.

\subsection{Constraints on energy injection}
\label{3.2}

In order for the exodus mechanism to set the relic density rather than the conventional freeze-out mechanism it is required that the number density of the relic DM $n_\zeta$ is entirely due to decays of $X$ states. It is important that the temperature of the relic sector remains lower than $m_\zeta$ such that the DM can not be thermally produced. Note that the requirement $T_{\rm relic}\ll m_\zeta<m_X$ implies that, for there to be initially a thermal distribution of $X$ states, the reheat temperature of the $X$ sector must be significantly higher than the relic sector, which could be due to preferential inflaton decay. Crucially, if the temperature of the relic sector is raised due to energy injection to the point that the exponential tail of the DM distribution is populated, then freeze-out of this thermally produced DM will likely dominate over the DM component produced via $X$ decays and thus ruin the exodus mechanism. If one assumes that the $\zeta$ have weak-scale self-interactions, then to avoid populating the tail of the distribution it is required that $T_{\rm relic}\lesssim m_{\zeta}/25$. Note that changes in the interaction strength will only lead to logarithmic deviations in the temperature bound. Additionally, a weaker requirement is that the three sectors should not equilibrate, in which case we expect a negligible  $X$ relic density which will disrupt the relationship $\Omega_{\rm DM}\approx5\Omega_B$. 

Consider the non-relativistic decays $X\rightarrow\zeta b$, for some final state $b$ carrying baryon number $B=1$. In order to satisfy the requirement that $T_{\rm relic}\lesssim m_{\zeta}/25$ it must be that the $X$ decay products have non-relativistic momenta and, working in this regime, we calculate the temperature of the relic sector after energy injection from the genesis sector $T^{(\infty)}_{\rm{relic}}$. The temperature of the $\zeta$ final states is related to the average kinetic energy
\beq
\frac{3}{2}k_BT^{(\infty)}_{\rm{relic}} = \langle {\rm KE}\rangle = \frac{\langle p_\zeta^2\rangle}{2m_\zeta}~.
\label{a1}
\eeq
We calculate $\langle p_\zeta^2\rangle$ working consistently in the non-relativistic approximation, from which the constraint $T^{(\infty)}_{\rm relic}\lesssim m_{\zeta}/25$ can be recast in terms of the masses
\beq
\frac{T^{(\infty)}_{\rm{relic}}}{m_\zeta}\simeq\frac{m_X^2-(m_b+m_\zeta)^2}{6m_\zeta^2}\lesssim\frac{1}{25}~.
\label{a2}
\eeq
In the simplest scenarios we expect that $m_{\zeta}\sim5~{\rm GeV}$ and $m_b\simeq1$ GeV in order to explain $\Omega_{\rm DM}\approx5\Omega_{B}$ and substituting values for $m_b$ and $m_\zeta$, the equality of eq.~(\ref{a2}) reduces to
\beq
\frac{T^{(\infty)}_{\rm{relic}}}{{\rm GeV}}\simeq\left(\frac{m_X}{6~{\rm GeV}}\right)^2-1~.
\eeq
For which the constraint $T^{(\infty)}_{\rm relic}\lesssim m_{\zeta}/25$ places an upper bound on
$m_X\lesssim6.5~{\rm GeV}$. In a wide range of scenarios there is sufficient freedom to choose the relative sizes of $m_\zeta$ and $m_X$ and thus viable models can be constructed.

\section{Exodus in the MSSM}
\label{Sec3}

Prior to considering more general implementations, we first present a simple supersymmetric realisation which offers a resolution to the well known problem of obtaining the correct relic abundance for (nearly) pure bino DM. Notably, the bino generally falls foul of relic density constraints since its annihilation cross section is p-wave suppressed. However, in the exodus mechanism the relic density is not set by the annihilation cross section and the bino can provide a viable DM candidate.  In this section we consider an implementation of the exodus framework in which the MSSM is appended with a genesis sector, we will not provide a full description of this sector, but we assume that it contains a SM singlet chiral superfield $\mb X$ which carries a conserved quantum number $\mathscr X=1$ with a particle-antiparticle asymmetry. 
 
If the inflaton decays preferentially to the genesis sector, it is conceivable that the reheat temperature of the genesis sector could be much higher than the visible sector $T^{\rm RH}_{\rm vis}\ll T^{\rm RH}_{\rm gen}$. If the reheat temperature of the visible sector is lower than the mass of the lightest supersymmetric particle then (initially) there will be no abundance of R-parity, $R_p$, odd states. Moreover, if the genesis sector is heated to temperature greater than the lightest $R_p$ odd state in the genesis sector, then visible sector superpartners can be generated through the decays from the genesis sector. This realises the exodus mechanism with the set of $R_p$ odd states identified with the relic sector and where the reheat temperatures of the relic and visible sectors are equal.  

Little is known about the reheat temperature of the universe after inflation apart from that it should be higher than the temperature of BBN (few MeV). In order to construct a viable model of bino DM via exodus production we require that the following hierarchy is respected
\beq
T^{\rm RH}_{\rm vis}\ll m_{\rm LSP} <m_X<T^{\rm RH}_{\rm gen}~.
\eeq
This ensures that the bino is not thermally produced (as discussed in Sect.~\ref{3.2}) and that there is initially a thermal abundance of $X$ states. In minimal models we expect that $m_{\rm LSP}\sim5~{\rm GeV}$ in order to explain $\Omega_{\rm LSP}\approx5\Omega_{B}$ and thus we require $T^{\rm RH}_{\rm vis}<m_{\rm LSP}/25\sim 200~{\rm MeV}$. For a 5 GeV neutralino LSP to be phenomenologically viable the state must be essentially purely bino in order to avoid direct production limits and invisible quarkonium decays \cite{quark}. Aspects of light bino phenomenology has been studied in  \cite{D'Eramo:2011ec,bino}.

The temperature of the early universe was certainly in excess of a few MeV at some stage, as a colder universe would alter the relative abundance of nucleons and deviations in these quantities are very restricted \cite{BBN}. There are two possible cosmological histories which could be realised in this model depending on the reheat temperature of the visible sector. In order to avoid observable deviations from BBN predications we require that either $T^{\rm RH}_{\rm vis}>$ few MeV or, if $T^{\rm RH}_{\rm vis}\lesssim {\rm MeV}$, that the energy injection due to $X$ decays must be sufficient to raise the temperature of the visible sector such that primordial nucleons are destroyed ($T_{\rm vis}\gtrsim100$ MeV), otherwise there is a risk that the experimentally verified ratios of nucleons may be disrupted. In the latter scenario, the requirement that energy injection heats the visible sector to $T^{\rm RH}_{\rm vis}\gg 1~{\rm MeV}$, combined with the condition that the energy injection must not lead to thermal bino production, results in a window for viable models and allows predictive scenarios to be constructed. Referring to the calculation of Sect.~\ref{3.2}, the temperature of the bino DM generated through $X$ decays for a 5 GeV bino LSP is parametrically $T^{(\infty)}_{\rm vis}\sim\left((m_X/6~{\rm GeV})^2-1\right)$ GeV and,  for an appropriate value of $m_X$, $X$ decays are sufficient to heat the visible sector to $T_{\rm vis}\gtrsim100$ MeV.

\begin{figure}[tb!]
\begin{center}
\includegraphics[height=45mm]{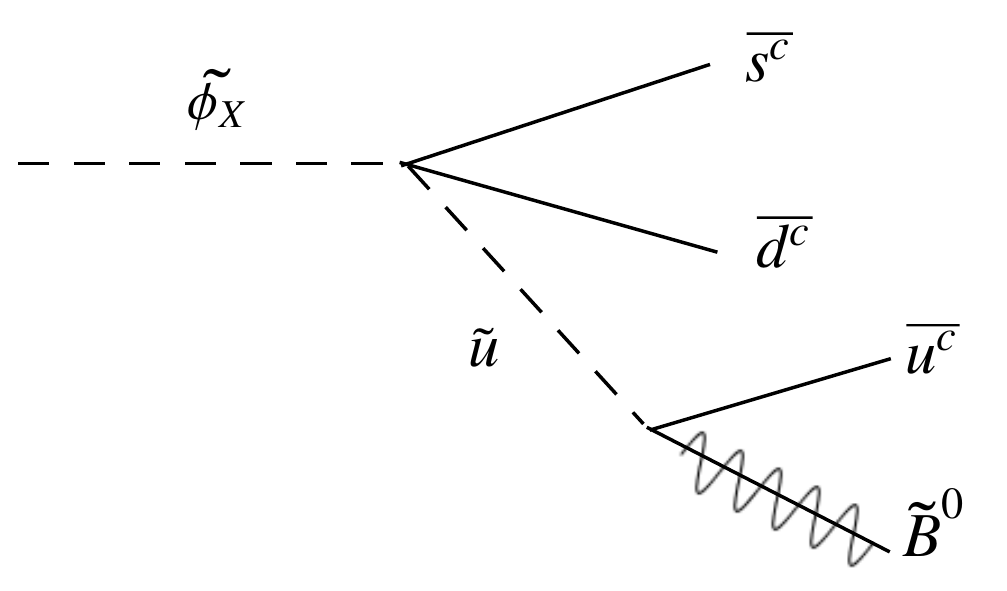}
\caption{A bino LSP provides good candidate for exodus dark matter.}
\label{Fig6}
\end{center}
\end{figure}

Additionally, it must be that the $X$ states generally decay well before the universe cools to $T\sim$MeV and that there are very few residual $X$ decays after this point, since injection of hadronic energy during BBN is greatly constrained \cite{BBN}. Suppose that the exodus mechanism proceeds via the portal operator $\epsilon_{jk}\mb{XU}_i\mb{D}_j\mb{D}_k$, where $i,j,k$ are generation indices. 
Under the generalised definition of $R$-parity, $R_p=(-1)^{2s+3(B-L+\mathscr X)}$, the $\mb X$ scalar (denoted $\phi_X$) is $R_p$ odd. The asymmetric component of the hidden sector state $\phi_X$ decays via a dimension five $B$-violating interaction producing an (off-shell) squark which subsequently decays to the bino LSP, as shown in  Fig.~\ref{Fig6}. Note that this is the lowest dimension $B$-violating trisector portal operator which can be constructed. The decay width due to this interaction is parametrically
 \beq
 \Gamma_X\sim\frac{\lambda^2g^\prime{}^2}{m^4_{\widetilde{q}}}\frac{\Delta^7}{512\pi^5}\frac{1}{M^2}~,
 \eeq
 where $\Delta\equiv m_X-m_{\rm LSP}-m_b$, the baryonic decay product has mass $m_b\simeq1$ GeV, and the operator  $\epsilon_{jk}\mb{XU}_i\mb{D}_j\mb{D}_k$ is dressed by the scale $M$ in the Lagrangian. Therefore typically the size for the decay width is 
 \beq
\Gamma_X\sim10^{-22}~{\rm GeV}  \left(\frac{\Delta}{5~{\rm GeV}}\right)^7 \left(\frac{1.5~{\rm TeV}}{m_{\widetilde{q}}}\right)^4\left(\frac{\lambda}{1}\right)^2 \left(\frac{10^4~{\rm GeV}}{M}\right)^2~,
 \eeq
which corresponds to a lifetime $\tau_X\sim10^{-2}$ s, with the indicated choices of parameters, and thus models is which the $X$ states decay before BBN can be constructed. Note that here the $B$-violating operator is suppressed by a scale $M\sim10$ TeV and therefore we do not expect collider constraints on this contact operator. Furthermore,  order of magnitude changes in scale $M$ can be accommodated given $\OO(1)$ changes in $\Delta$.

\section{Generalised models of  exodus}
\label{Sec4}

There are a myriad of implementations for the general exodus framework,  we shall give a flavour of the possibilities and demonstrate that interesting models can be constructed.

\subsection{B-violating exodus}
\label{Sec4.1}

First we consider generalisations of Sect.~\ref{Sec3}, in which the baryon asymmetry and the DM relic density are cogenenerated via exodus. The most straightforward manner in which to achieve this is to assume that the $X$ decay to states carrying baryon number in the visible sector. In the SM the lowest dimension gauge invariant $B$-violating operator is $u^cd^cd^c$  and  a possible trisector portal operator is $\frac{1}{M^3}X\zeta u^cd^cd^c$ where $X$ is a complex scalar or Dirac fermion carrying $\mathscr X=1$ and  the DM $\zeta$ is a Majorana fermion or real scalar. This operator conserves $B-L+\mathscr X$, whilst violating $B$ and $\mathscr X$ separately. Decays via this operator produce anti-neutrons $X\rightarrow \zeta \bar n^0$ and it is appropriate to rewrite the operator as a normalisable term in the effective low energy Lagrangian 
\begin{equation}
\mathcal{L}_{\rm eff}\supset
\lambda\left(\frac{\Lambda_{\rm QCD}}{M}\right)^3 X\zeta \bar n^0~,
\end{equation}
where $\Lambda_{\rm QCD}\simeq200~{\rm MeV}$ is the QCD scale and $\lambda$ is a dimensionless constant.

As discussed in the previous section, there are strong bounds on hadronic decays during BBN and thus first we wish to ascertain the parameter range in which the $X$ states decay well before BBN.  In minimal models in which $X$ has charge $\mathscr X=1$, to explain $\Omega_\zeta/\Omega_B\approx5$ we require that the DM mass is $m_\zeta\sim5~\GeV$. If the genesis sector state carries large (fractional) $\mathscr X$ number then the mass of the DM state $\zeta$ can be raised (lowered) and still account for the coincidence of relic densities. For simplicity, and to demonstrate that viable models can be constructed, let us assume that the genesis sector and the visible sector maintain roughly the same temperature.\footnote{If the genesis sector is cooler than the visible sector, the $X$ states are Boltzmann suppressed earlier and their number density is depleted more quickly, which leads to a larger allowed parameter space.}  
In this case the condition that $X\bar X$ annihilations freeze-out after the abundance is set by the asymmetry is encapsulated in Fig.~\ref{FigC}, observe that for $m_X\sim5-10~\GeV$ the freeze-out temperature is required to be lower than $T_{\rm gen}^{(\rm C)}\lesssim250~\MeV$. Comparing with the freeze-out temperature, given in eq.~(\ref{FOT}),
\begin{equation}
T^{(\rm{FO})}_{\rm gen}\sim10^{-4}~\GeV~\left(\frac{1~\GeV^{-2}}{\sigma}\right)^{1/2}\left(\frac{m_X}{10~\GeV}\right)^{1/2}~,
\label{exFO}
\end{equation}
and we note that for an appropriate value of the $X\bar X$ annihilation cross section that freeze-out will occur well after the symmetric component of $X$ has annihilated.
To ensure that the $X$ states decay before BBN, but after the symmetric component has annihilated, its lifetime must lie in the range $10^{-5}~{\rm s}<\tau_X<1~{\rm s}$, where we have used eq.~(\ref{tmin}) to convert between temperature and time.
The lifetime for $X$ decaying via $X\rightarrow\zeta \bar n^0$, as given in eq.~(\ref{tau}), is
\begin{equation}
\tau_X\sim 10^{-2}s\left(\frac{10^{-2}}{\lambda}\right)^2 \left(\frac{10\, \mathrm{GeV}}{m_{ {X}}}\right)\left(\frac{10^5}{\Lambda_{\rm{QCD}}/M}\right)^6~.
\end{equation}
Thus viable models of $\zeta$ DM using the portal operator $X\zeta u^cd^cd^c$ can be constructed.

\subsection{L-violating exodus}
\label{Sec4.2}

\begin{figure}[t!]
\begin{center}
\includegraphics[height=40mm]{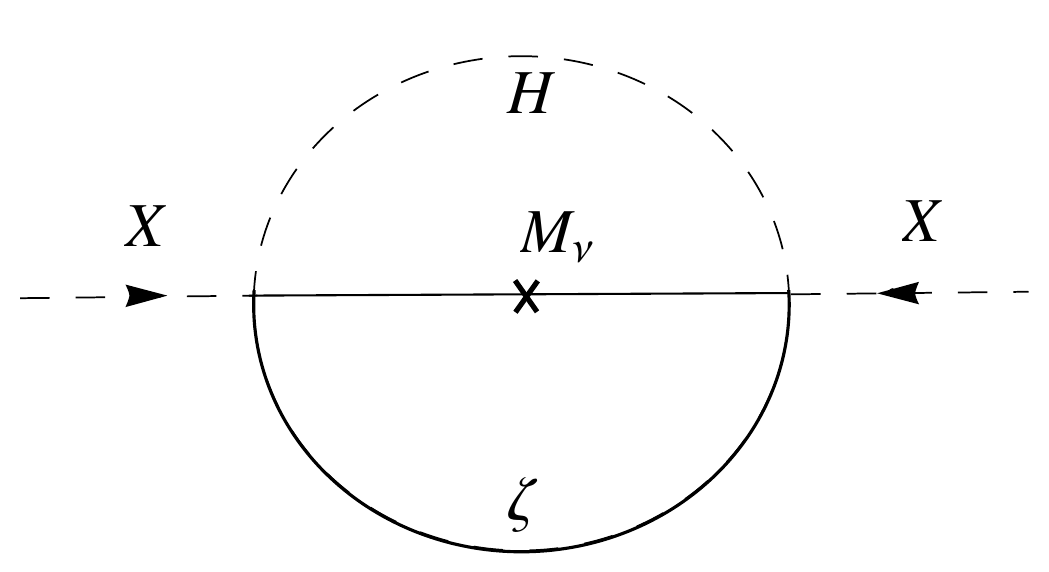}
\vspace{5mm}
\caption{Radiatively generated $\mathscr{X}$-violating mass for $X$ due to Majorana neutrino mass insertion. \label{FigX} }
\end{center}
\end{figure}

An alternative approach to using a hidden sector particle asymmetry to directly generate the baryon asymmetry, is to first induce an asymmetry in lepton number above the EWPT and  subsequently, sphaleron processes will transfer this asymmetry to baryon number \cite{BviaL}. The SM singlet, $R_p$-violating portal operator $\mb{X\zeta LH}$ which conserves $B-L+\mathscr X$ provides a possible candidate for the trisector operator. The exodus mechanism can then be implemented via the following process $X\rightarrow\bar{\nu} h \zeta$ (or similar). 
Similar to the $B$-violating exodus scenario, the $X\bar X$ annihilation cross section must be sufficiently large that the symmetric component is depleted before $X\bar X$ annihilations freeze-out and also before the $X$ states decay. Moreover, for this sharing to be successful decays of $X$ to leptons must occur prior to the EWPT and thus requires that $\tau_X\lesssim10^{-10}~\rm{s}$. This can result in tension between the requirement for $X$ to decay promptly and yet be sufficiently long-lived that the symmetric component of $X$ is adequately depleted. 
In particular, if the visible sector and genesis sector have comparable temperatures then, by comparison with Fig.~\ref{FigC}, in order for the symmetric component to be removed before the EWPT ($T_{\rm{vis}}\sim100~\GeV$) the $X$ states must have $m_X\gtrsim 5$ TeV. We compare this to the lifetime of $X$ as given by eq.~(\ref{tau}) 
\begin{equation}
\tau_X\simeq2\times10^{-11}~{\rm s}~\left(\frac{10^{-8}}{\lambda}\right)^2 \left(\frac{5\, \mathrm{TeV}}{m_{ {X}}}\right)~.
\end{equation}
However, in order to avoid large heating of the relic sector due to $X$ decays, the $\zeta$ states must also generally have TeV scale masses. However, if  $T_{\rm gen}\ll T_{\rm vis}$, then the $\zeta$ and $X$ can have GeV masses, as the symmetric component of $X$ will become depleted earlier.

Furthermore, the presence of $L$-violating operators, such as Majorana neutrino masses, can generate an effective symmetry-violating Majorana mass for $X$ which is enhanced due to the non-renormalisable portal operator.
In this case care must be taken as this can lead to fast $X$-$\bar X$ oscillations which will erase the asymmetry in the genesis sector and ruin the mechanism. 
To illustrate this point, we consider the simple example of the dimension five operator $\frac{1}{M}X\zeta lh$ (where $M$ is the mass of the heavy mediator state which is integrated out), in the presence of a Majorana neutrino mass $M_\nu\nu_L\nu_L$.  As well as a symmetry-preserving Dirac mass $m_X$  for $X$, there is also a radiatively generated $\mathscr{X}$-violating Majorana mass $M_X$, feeding-in from  the Majorana neutrino mass, as shown in Fig.~\ref{FigX}. The size of the symmetry violating mass is parametrically
\beq
M_X^2\simeq\frac{1}{(16\pi^2)^2}\frac{\Lambda^3M_\nu}{M^2}
\simeq\left(2\times10^{-3}\right)^2\left(\frac{\Lambda}{10^6~{\rm GeV}}\right)^3\left(\frac{M_\nu}{10^{-11}~{\rm GeV}}\right)\left(\frac{10^4~{\rm GeV}}{M}\right)^2
\eeq
where $\Lambda$ is some UV-cutoff at which the Feynman diagram of Fig.~\ref{FigX} breaks-down.
In order for the exodus mechanism to be successful, it must be that the symmetry violating $X$-mass be sufficiently small $M_X\ll m_X$, such that oscillations do not erase the $\mathscr{X}$ asymmetry before it is transferred to the visible sector, and subsequently to the baryons. The effects of DM-antiDM oscillations in models of asymmetric DM have been studied in \cite{oscillation}.
This is not a problem if $\Lambda$ and $M$ are comparable and relatively low ($\lesssim10^9$ GeV), however GUT scale values can not be accommodated.
This issue can be avoided if lepton number is not violated in generating the neutrino masses, with the neutrinos instead acquiring small symmetry preserving Dirac masses, see for e.g.~\cite{Dirac}. Thus, while $L$-violating models are possible, they are typically more constrained and less appealing than the $B$-violating scenario.

\subsection{Asymmetric dark matter via exodus}

By changing the $\mathscr X$ charges of the states involved in the trisector interaction both symmetric and asymmetric DM can result from the exodus mechanism. The operators studied thus far have had the property that they violate $B$ or $L$ by 1 unit. If we suppose that the state $X$ carries $\mathscr X=Q$ then to conserve $B-L+\mathscr X$ it follows that $\zeta$ must carry $\mathscr X=Q-1$. In this case the states $\zeta$ inherit the asymmetry of the genesis sector identically to the baryons. It is understood that $\zeta$ must be a Dirac fermion or complex scalar in order to carry $\mathscr X$ number. Both the baryon-violating and lepton-violating scenarios discussed above can be recast in terms of asymmetric DM in this manner. For example, suppose that the DM relic state $\zeta$  is a  complex scalar (or Dirac fermion) with $\mathscr X=1$, and the state $X$ carries $\mathscr X=2$, if the connector operator is $X\zeta u^cd^cd^c$, which violates $B$ by 1 unit, then the decays of $X$ generate a population of $\zeta$ particles, whilst the number density of $\bar\zeta$ remains zero. As in asymmetric DM models \cite{ADM}, the final DM relic density is composed of only $\zeta$ states, with no abundance of $\bar\zeta$.

\section{Concluding remarks}
\label{Sec6}

We have outlined a new framework for explaining the observed relation $\Omega_{\rm{DM}}\approx5\Omega_B$. The exodus scenario assumes that the number density of the DM state is initially near-zero and is only generated due to decays from a second hidden sector  possessing a particle-antiparticle asymmetry. This produces DM and generates a $B-L$ asymmetry, resulting finally in the observed baryon asymmetry. In contrast to models of asymmetric DM, the DM can be a real scalar or Majorana fermion, thus presenting distinct phenomenology.\footnote{Although, see  \cite{Okada:2012rm} for an alternative proposal for obtaining Majorana DM in an asymmetric DM setting.} In particular, models of asymmetric DM can not typically have annihilation signals, however, if the exodus DM state $\zeta$ is a real scalar or Majorana fermion DM annihilations can occur, and thus potentially produce observable indirect detection signals with annihilation profiles.

Furthermore, in models of asymmetric DM the symmetric component of the DM must annihilate in order for the asymmetry to set the relic density and this leads to strong constraints. These limits rule out a large class of models if the symmetric component annihilates directly to the visible sector \cite{MarchRussell:2012hi} and there are also constraints on annihilations to light hidden sector states from DM self-interactions \cite{Lin:2011gj}. In contradistinction, the genesis sector states $X$ of exodus models, which possess a particle-antiparticle asymmetry, can have large annihilation cross sections to the visible sector or to light hidden sector states whilst avoiding these bounds, since the DM relic density is not composed of the $X$ states, but rather the relic sector states $\zeta$ (which do not necessarily carry an asymmetry). One potential observable of annihilations to light hidden sector states is that this could increase the effective number of neutrino species, similar to as studied in \cite{Blennow:2012de}. 

Further, in the supersymmetric setting we argued that if the reheat temperature is lower than the mass of the LSP then the $R_p$ odd states can play the role of the relic sector and a bino LSP provides a good candidate for DM in the exodus framework. This is particularly interesting since the correct DM relic density can not generally be obtained for a bino LSP via conventional freeze-out as its annihilation cross section is too small. This new framework opens many new possibilities for DM model building and begins to explore the prospect of multiple sector DM phenomenology.

\section*{Acknowledgements}
I am grateful to  Pavel Fileviez P\'erez, Edward Hardy, Ulrich Haisch, Matthew McCullough, Stephen West, and especially, John March-Russell for useful comments and discussions. This work was supported by a Charterhouse European Bursary and an award from the Vice-Chancellors Fund, University of Oxford.


\appendix

\section{Boltzmann equations for exodus}
\label{App}

The change in the $\mathscr{X}$ asymmetry $n_{\mathscr X}\equiv n_X-n_{\overline{X}}$ can be described by the Boltzmann equations (see e.g.~\cite{standard}). To first order in the small trisector coupling $\lambda$, assuming two-body $X$ decays to DM states $\zeta$ and baryons $b$, this can be expressed as
\begin{equation}
\begin{aligned}
\dot n_{\mathscr X} + 3Hn_{\mathscr X} 
&=\int {\rm d}\Pi_X {\rm d}\Pi_b {\rm d}\Pi_\zeta (2\pi)^4\delta^{(4)}(p_X-p_b-p_\zeta)
\left[|M|^2_{b\zeta\rightarrow X} f_bf_\zeta-|M|^2_{X\rightarrow b\zeta} f_X\right]\\
&~-\int {\rm d}\Pi_X {\rm d}\Pi_b {\rm d}\Pi_\zeta (2\pi)^4\delta^{(4)}(p_X-p_b-p_\zeta)
\left[|M|^2_{\overline{b\zeta}\rightarrow \overline{X}} f_{\overline{b}}f_{\overline{\zeta}}-|M|^2_{\overline{X}\rightarrow \overline{b\zeta}} f_{\overline{X}}\right]
\label{BE}
\end{aligned}
\end{equation}
where we have neglecting statistical factors. Note, ${\rm d}\Pi_i=\frac{{\rm d}^3p_i}{(2\pi)^32E_i}$ and the phase space distribution function $f_i$ is related to the number density by
\beq
n_i=\frac{g_i}{(2\pi)^3}\int {\rm d}^3p f_i~,
\eeq
where $g_i$ is the number of internal spin degrees of freedom. For a state in thermal equilibrium at temperature $T$ the phase space distribution function is of the form
\beq
f_i\simeq \exp\left(-\frac{E_i}{T}\pm\frac{\mu_i}{T}\right)~,
\eeq
where $\mu_i$ is the chemical potential which describes unbalances between particle-antiparticle number densities due to asymmetries: $\eta_{i}\equiv (n_i-n_{\overline{i}})/S\propto\mu_i/T$. This term takes opposite signs for particles and antiparticles and is absent in the case of self-adjoint fields. 

Immediately prior to $X$ decays the number density of relic DM is negligible $n_{\zeta,\overline{\zeta}}\approx0$ and it follows $f_\zeta\approx f_{\overline{\zeta}}\approx0$. Moreover, as the processes do not feature large CP violation, $X$ and $\overline{X}$ processes can be treated equally, hence $|M|^2_{b\zeta\rightarrow X}=|M|^2_{\overline{b\zeta}\rightarrow \overline{X}}$ and by CPT invariance $|M|^2_{b\zeta\rightarrow X}=|M|^2_{ \overline{X}\rightarrow\overline{b\zeta}}$. Thus eq.~(\ref{BE}) reduces to
\begin{equation}
\begin{aligned}
\dot n_{\mathscr X} + 3Hn_{\mathscr X} &\simeq\int {\rm d}\Pi_X 2m_X\Gamma_X \left(f_{\overline{X}}-f_X\right)~,
\label{XQ}
\end{aligned}
\end{equation}
where we have identified the decay width $\Gamma_X=\frac{1}{2m_X}\int\Pi_b {\rm d}\Pi_\zeta(2\pi)^4\delta^{(4)}(p_X-p_b-p_\zeta)|M|^2_{X\rightarrow b\zeta} $. Converting the integration variable from momentum to energy (following \cite{Hall:2009bx}) we obtain
\begin{equation}
\begin{aligned}
\dot n_{\mathscr X} + 3Hn_{\mathscr X} 
& \simeq \int_{m_X}^\infty \frac{{\rm d}E_{X}}{2\pi^2} m_X\Gamma_X  \sqrt{E_X^2-m_X^2} e^{-E_X/T}\left(e^{\mu_X/T}-e^{-\mu_X/T}\right)\\
& \simeq \frac{m_X^2 \Gamma_XT}{2\pi^2}  K_1\left(\frac{m_X}{T}\right) \left(e^{\mu_X/T}-e^{-\mu_X/T}\right)~,
\end{aligned}
\end{equation}
where $K_i$ a the modified Bessel function of the second kind. 
 Re-expressing this in terms of $\Delta_{\mathscr{X}}$, and using the asymptotic form of the Bessel function $K_1(x)\sim \sqrt{\frac{\pi T}{2m_X}} \exp\left(-\frac{m_X-\mu}{T}\right)$ gives
\begin{equation}
\begin{aligned}
\dot \Delta_{\mathscr{X}}
&\simeq  \frac{ \Gamma_X}{S} \left(\frac{m_X T}{2\pi}\right)^{3/2}  \left(e^{-(m_X+\mu_X)/T}-e^{-(m_X-\mu_X)/T}\right)
\simeq - \Gamma_X \Delta_{\mathscr{X}}~,
\label{QW}
\end{aligned}
\end{equation}
where we collected terms into the number densities, as stated in eq.~(\ref{n}).
It follows that 
\begin{equation}
\int\frac{{\rm d} \Delta_{\mathscr{X}}}{ \Delta_{\mathscr{X}}}\simeq - \int {\rm d}t~ \Gamma_X\simeq \Gamma_X\int{\rm d}T ~\frac{1}{HT}~,
\end{equation}
and integrating between early time and late time, at temperature $T$, yields
\begin{equation}
\begin{aligned}
\log\left(\frac{\Delta_{\mathscr{X}}^{\rm (init)}}{\Delta_{\mathscr{X}}(T)}\right)\simeq - ~\frac{\Gamma_X}{2}\frac{\MP}{1.66\sqrt{g_*^{\rho}}}\left(\frac{1}{T^2_{\rm UV}}-\frac{1}{T^2}\right)~.
 \end{aligned}
\end{equation}
where $\Delta_{\mathscr{X}}^{\rm (init)}$ is the initial $\mathscr{X}$ asymmetry and $\Delta_{\mathscr{X}}(T)$ is the asymmetry at temperature $T$. The quantity $T^2_{\rm UV}$ is the temperature at which the primordial $\mathscr{X}$ asymmetry is generated and which in certain cases may be taken to be the reheat temperature. 

For the asymmetry to be transferred we require that $\Delta_{\mathscr{X}}(T)\ll {\Delta}_{\mathscr{X}}^{\rm init}$, thus
\begin{equation}
\begin{aligned}
\Delta_{\mathscr{X}}(T)
&\sim {\Delta}_{\mathscr{X}}^{\rm init}\times10^{-9}\exp\left[2\left(1- \left(\frac{{\rm GeV}}{T}\right)^2\left(\frac{\Gamma_X}{10^{-18}~{\rm GeV}}\right)\left(\frac{10}{\sqrt{g_*^{\rho}}}\right)\right)\right]~.
 \end{aligned}
\end{equation}
For instance, for the asymmetry to be transferred via $X$ decays prior to $T\sim$ GeV (assuming that the visible and hidden sectors maintain comparable temperatures) the width must be $\Gamma\sim10^{-18}$ GeV and, hence, the lifetime of $X$ is $\tau\sim10^{-6} ~{\rm s}$.  Conversely, to ensure that the asymmetry is transferred before BBN at $T\sim$MeV one requires $\Gamma\sim10^{-24}$ GeV, hence $\tau\sim1~s$, as expected.
In the case that the temperatures between the sectors undergo different thermal evolutions then $R\equiv T_{\rm vis}/T_{\rm gen}\neq1$ and the above equation will depend on this quantity
\begin{equation}
\begin{aligned}
\Delta_{\mathscr{X}}(T)
&\sim {\Delta}_{\mathscr{X}}^{\rm init}\times10^{-9}\exp\left[2\left(1- \left(\frac{{\rm GeV}}{T_{\rm vis}}\right)^2\left(\frac{R}{100}\right)^2\left(\frac{\Gamma_X}{10^{-22}~{\rm GeV}}\right)\left(\frac{10}{\sqrt{g_*^{\rho}}}\right)\right)\right].
 \end{aligned}
\end{equation}

Further, by conservation of the combination of quantum number $B-L+X$ it follows that $\Delta_{\mathscr{X}}\propto-\Delta_B\propto-\Delta_\zeta$. Moreover, for two body $X$ decays or in the case that $\zeta$ does not carry a conserved charge (e.g.~for self-adjoint $\zeta$) then $\Delta_{\mathscr{X}}=-\Delta_B$, which is clear from analysis of the corresponding Boltzmann equation 
\begin{equation}
\begin{aligned}
\dot n_{ B} + 3Hn_{ B} 
&=\int {\rm d}\Pi_X {\rm d}\Pi_b {\rm d}\Pi_\zeta (2\pi)^4\delta^{(4)}(p_X-p_b-p_\zeta)
\left[|M|^2_{X\rightarrow b\zeta} f_X-|M|^2_{b\zeta\rightarrow X} f_b f_\zeta\right]\\
&~-\int {\rm d}\Pi_X {\rm d}\Pi_b {\rm d}\Pi_\zeta (2\pi)^4\delta^{(4)}(p_X-p_b-p_\zeta)
\left[|M|^2_{\overline{X}\rightarrow\overline{b\zeta}} f_{\overline{X}}-|M|^2_{\overline{b\zeta}\rightarrow \overline{X}} f_{\overline{b}}f_{\overline{\zeta}}\right]~.
\end{aligned}
\end{equation}
Then, similarly to previously, as $f_\zeta\approx f_{\overline{\zeta}}\approx0$ this reduces to
\begin{equation}
\begin{aligned}
\dot \Delta_{B} &\simeq\frac{1}{S}\int {\rm d}\Pi_X 2m_X\Gamma_X \left(f_X-f_{\overline{X}}\right)\simeq-\dot \Delta_{\mathscr{X}}~,
\end{aligned}
\end{equation}
where the final equality can be seen by comparison to eq.~(\ref{XQ}). A similar argument can be made for the $\mathscr{X}$ asymmetry inherited by the $\zeta$ states in the case that they carry $\mathscr{X}$ number. On the other hand, in the case that the relic DM $\zeta$ is self-adjoint then the Boltzmann equation which describes their evolution is
\begin{equation}
\begin{aligned}
\dot n_{\zeta} + 3Hn_{\zeta} 
&=\int {\rm d}\Pi_X {\rm d}\Pi_b {\rm d}\Pi_\zeta (2\pi)^4\delta^{(4)}(p_X-p_b-p_\zeta)
\left[|M|^2_{X\rightarrow b\zeta} f_X-|M|^2_{b\zeta\rightarrow X} f_b f_\zeta\right]\\
&~+\int {\rm d}\Pi_X {\rm d}\Pi_b {\rm d}\Pi_\zeta (2\pi)^4\delta^{(4)}(p_X-p_b-p_\zeta)
\left[|M|^2_{\overline{X}\rightarrow\overline{b}\zeta} f_{\overline{X}}-|M|^2_{\overline{b}\zeta\rightarrow \overline{X}} f_{\overline{b}}f_{{\zeta}}\right].
\end{aligned}
\end{equation}
Thus, following analogous steps to (\ref{XQ})-(\ref{QW}), this implies
\begin{equation}
\begin{aligned}
\dot Y_{\zeta}
&\simeq \Gamma_X~\left(\frac{n_X+n_{\overline{X}}}{S} \right)
\simeq \Gamma_X\Delta_{\mathscr{X}}\simeq -\dot\Delta_{\mathscr{X}}~,
\end{aligned}
\end{equation}
where in the intermediate step we have written $n_X+n_{\overline{X}}=S\Delta_{\mathscr{X}}+2n_{\overline{X}}$ and used that $n_{\overline{X}}\approx0$. These results conform with our expectations, as expressed in the main body of the paper and as depicted in Fig.~\ref{Fig2}. Note that washout effects enter only at order $\lambda^2$ and higher in the feeble trisector coupling and thus can  be neglected.

\end{document}